\documentclass[pdftex,moreauthors]
{article}

\usepackage{soul}
\usepackage{amsmath}

\begin{document}

\huge{
Primordial Perturbations Including Second-Order Derivatives of the Inflationary Potential}


\vskip 1cm
\large
{{Paulo Custodio} 
 $^{1,}$*, {Cristian R. Ghezzi} 
 $^{2}$, {Nadja S. Magalhaes} 
 $^{3}$ and Carlos Frajuca $^{4}$
 \vskip 1cm





\normalsize
$^{1}$ \quad Instituto de Ciências Exatas e Tecnologia, Universidade Paulista, Av. Marques de São Vicente, 3001, Bairro Água Branca,  Distrito Barra Funda, Zona Oeste, Sao Paulo
  { 05036-040}
,  
 Brazil\\
$^{2}$ \quad Independent Researcher, Bahia Blanca 
{ 8000}, {Provincia de Buenos Aires},  {Argentina}
; gluon00@yahoo.com\\
$^{3}$ \quad Physics Department, Federal University of S{ã}o Paulo, Rua Sao Nicolau, 210, Diadema 09913-030,  Brazil; nadja.magalhaes@unifesp.br\\
$^{4}$ \quad Mechanics Department, Instituto Federal de São Paulo,   Rua Pedro Vicente, 625, Sao Paulo 01109029,   Brazil; {frajuca@gmail.com} 

{Correspondence: paulo.custodio@docente.unip.br}




\abstract{In inflationary cosmology, the form of the potential is still an open problem. In this work, second-order effects of the inflationary potential are evaluated and related to the known formula for the primordial perturbations at a wide range of scales. We found effects that may help to unravel the unknown inflationary potential form and impose new constraints on the parameters that define this potential.
In particular, we demonstrate that even slight deviations in the inflationary potential can lead to significant differences in the calculated spectra if inflation persists sufficiently long and the normal modes of perturbations are affected by these variations.
}

\section{Introduction}
Inflationary models have been instrumental in resolving several critical puzzles of the original Big Bang theory, including the horizon, flatness, and the generation of primordial fluctuations necessary for structure formation~\cite{Guth,starobinski82}. Initially inspired by grand unified theories in particle physics, 
these models propose a period of rapid exponential expansion of the Universe, known as inflation, which stretches quantum fluctuations to astrophysical scales~\cite{starobinski82,HawkingMoss,LiddleLyth,Cosmo}.

The quantum mechanical fluctuations of a scalar field called the inflaton give rise to perturbations that evolve under the influence of a specific potential. During inflation, these perturbations start at sub-Hubble scales, expand exponentially during the slow-roll inflation period, and eventually exit the Hubble horizon, laying the groundwork for forming cosmic structures observed today. Subsequently, as inflation ends, the inflation field ``reheats'', generating the hot primordial soup of the Big Bang while losing its energy~\cite{Kofman,Liddle2003}.

Following inflation, the Universe's evolution transitions to the classical Friedmann--Robertson--Walker (FRW) cosmological model, further refined to include dark energy in the $\Lambda$-CDM (cold dark matter plus cosmological constant) model~\cite{LiddleLyth}.
The quantum fluctuations during inflation result in slight variations in the Universe's mean density, known as primordial density perturbations. These fluctuations are imprinted in the cosmic microwave background radiation (CMB) and manifest as perturbations in its intensity and~polarization. 

Observations of the CMB, notably by satellites such as COBE, WMAP, and PLANCK, have provided crucial insights into the Universe's early evolution~\cite{Planck2018params, Planck2018inflation, ade2016, hinshaw2013}. 
The Sachs--Wolf effect, for instance, allows the detection of perturbations in the CMB radiation, which are indicative of density fluctuations in the early Universe. 

Numerical simulations have demonstrated that the density perturbations' predicted spectrum and amplitude are consistent with the observed cosmic structures formed over the Universe's 14-billion-year history. Consequently, observations of the CMB, large-scale structure formation, and nucleosynthesis serve as constraints for theoretical models, particularly inflationary and Big Bang models.

Despite the successes of inflationary models, specific unresolved issues persist in cosmology, such as the transition of quantum fluctuations to classical overdensities. This transition from a symmetric quantum state to a classical non-symmetric one poses a fundamental challenge in physics~\cite{okon}. Understanding this transition is crucial for accurately reconstructing the inflationary potential, as it influences parameters such as reheating temperature, primordial particle production, and perturbation amplitudes~\cite{lidsey97,dalianis2024}.

The most critical issue in current cosmology is the tension between early- and late-time determinations of the Hubble constant, $H_0$, with values derived from Planck CMB measurements differing from those obtained via redshift measurements of type Ia supernovae, as seen in the SH0ES project~\cite{riess}. Reference~\cite{divalentino} thoroughly catalogs various approaches to addressing the Hubble tension, exploring an array of models, such as early and late dark energy, dark energy models with multiple degrees of freedom, extensions involving extra relativistic components, models with additional interactions, unified cosmologies, modifications to gravity, inflationary models, altered recombination histories, critical phenomena physics, and other alternative propositions.

Some researchers have considered the second derivative of the potential in different inflationary models. Bellini~\cite{bellini} considered a non-minimally coupled inflationary model, incorporating the second derivative of the potential to determine constraints for the model. Lin {\it et al.}~\cite{lin} ruled out the natural inflation model based on observational constraints within the single-field inflationary model. Chiba~\cite{chiba} related the inflationary potential to the spectral index and its dependence on the e-folding number. Belfiglio and Luongo~\cite{Belfiglio24} proposed a non-minimally coupled quintessence model that leads to a cosmological phase transition, particle production, and density perturbations. In this model, the vacuum energy decays into particles beyond baryons, offering a potential solution to the cosmological constant problem (see also~\cite{Luongo23}).

This paper contributes to this investigation by incorporating second-order corrections to the inflationary potential when calculating quantum fluctuations. These corrections, often disregarded in past studies due to stringent slow-roll conditions~\cite{NewInflation,InflaQuantumCosmo,liddle94}, are examined in light of the recent findings. Although we do not aim to resolve the Hubble tension, our approach explores potential solutions within single-scalar-field inflation, specifically by relaxing slow-roll conditions near the end of inflation. The traditional slow-roll conditions, namely, $|{V'(\phi)/V(\phi)}| \ll \sqrt{24 \,\pi\, G}$ and $|V''(\phi)/V(\phi)| \ll \sqrt{24\, \pi \,G}$, suffice for sustaining inflation but may not be strictly necessary and could be adjusted as inflation concludes ({see}
~\cite{weinberg}, p. 211). We posit that the potential shape can meaningfully impact measured values of the spectral index $n_s$, e-folding number $N$, and tensor-to-scalar ratio $r$. We use the CMB density contrast ($\delta \rho/\rho$), e-folding number, and spectral index as constraints. While the tensor-to-scalar ratio primarily provides upper bounds, the rise of gravitational wave observatories and antennas provide promise for its use for further constraining models. Some of us (CF and NSM) are part of the team that developed the Mario Schenberg spherical gravitational wave antenna and are concerned with its application to cosmology-related problems~\cite{Schenberg}.

In this study, we add the second derivative of the inflationary potential into normal modes, revealing a non-trivial dependence between Fourier modes and the shape of the potential. We demonstrate that even slight deviations from the slow roll can lead to significant differences in the calculated spectra if inflation persists sufficiently long.

\section{Inflation}
Inflation was a brief epoch of exponential expansion of the Universe.
During this period, the Universe  accelerated as $a(t)=e^{Ht}\,
$; the expansion was superluminal and preceded the hot Big Bang~\cite{Cosmo2}.
Here, $a(t)$ stands for the scale factor of the Universe, $H=H(\varphi)$ is the Hubble rate, and $t$ is the cosmic time. A scalar field, dominated by its potential energy  $V(\varphi)$, drives inflation. This potential dominates all forms of energy of that period, and the pressure of this field must be negative. 

The field that drives inflation,  $\varphi$, is a single scalar field. There are some models including two fields (or more) called hybrid inflation.
Although this field is roughly uniform, the Heisenberg uncertainty principle predicts quantum fluctuations.
Therefore, some regions have different scalar field $\varphi$ values and inflate at slightly different rates. 
This phenomenon triggers perturbations in the background metric.

To understand how the quantum perturbations arise,  consider the cosmological horizon in a quasi-de Sitter spacetime, which is the distance ($R$) that light can cross in comoving coordinates since the beginning of the expansion:
\begin{equation}
\label{eq1}
R=\int_{t_{e}}^{t_{0}}{\frac{cdt}{a(t)}}=\int_{0}^{r_{e}}{\frac{dr}{\sqrt{1-Kr^{2}}}}\,=c/H\,,
\end{equation}
where {$t_e$} 
 and {$t_0$} are the times of the beginning and the end of inflationary expansion, respectively, while $r_{e}$ yields the size of this patch from zero to when inflation finishes.

The presence of this horizon produces thermal radiation, similar to Hawking radiation in a black hole, but with the black hole region turned inside out.  The cosmic horizon restricts the modes of the zero-point quantum fluctuations like in the Casimir effect~\cite{casimir48}. Consequently, the Heisenberg 
uncertainty principle implies quantum fluctuations in the momentum of the field, which satisfies the quantum inequality~\cite{InflaQuantumCosmo}
\begin{equation}
\label{eq2}
\Delta{p}\geq{{{\hbar}\frac{H}{c}}}\,,
\end{equation}
resulting in a low cut-off for the momentum (it is the direct application of the uncertainty principle into Equation (\ref{eq1})).

{On the other hand, the cosmic horizon has a Hawking temperature~\cite{HawkingMoss, InflaQuantumCosmo} given by
%
\begin{equation}
\label{eq3}
\delta\varphi=k_{B}T={\frac{{\hbar}H}{2\pi}}=\frac{H}{2\pi}\,.
\end{equation}

{From now on, we consider the natural system of units, in which $c=\hbar=k_{B}=1$ and the gravitational constant is $G={1}/{M_{pl}^{2}}$, implying $T=H/{2\pi}$,
with \mbox{$M_{pl}=1.22{~\times~}10^{19}\,$GeV;}} 
the Planck time and Planck length are given by $T_{pl}=0.53{~\times~}10^{-45}\,$s and $L_{pl}=1/M_{pl}=1.6{~\times~}10^{-33}\,$cm, respectively. 
The Hubble parameter has energy units $[H]=GeV$, and the energy density of 
the scalar field units is  $GeV^{4}$}.

The quantum fluctuations created during inflation stretch out of the causal horizon while the horizon remains stationary~\cite{InflaQuantumCosmo,Cosmo2}. The fluctuation becomes frozen as it crosses the horizon and, presumably, converts to a classical perturbation.

After inflation, the cosmic expansion continues at subluminal velocities, and the fluctuations return to the causal horizon. Therefore, the most essential perturbations for structure formation arise from the fluctuations excited
near the end of inflation~\cite{Cosmo2}.
The squared curvature perturbation
$\delta^{2}_{H}(k)$ becomes frozen when it crosses the horizon. This perturbation corresponds to $k=aH$, where $k$ is the comoving wave number of a perturbation's mode.

The curvature perturbation
$\delta_{H}(k)$ is related to the Friedmann equations, 
which are given by~\cite{InflaQuantumCosmo}
\begin{eqnarray}
\label{eq4}
&&{H^{2}(\varphi)}={\frac{8\pi}{3{M_{pl}}^{2}}}{V(\varphi)}\,, \\ 
\label{eq5}
&&{\frac{\ddot{a}}{a}}=-{\frac{4\pi}{3{M_{pl}}^{2}}}{(\rho+3P)}\,.
\end{eqnarray}
{The} 
 equation of state for the inflationary field is~\cite{InflaQuantumCosmo}
\begin{equation}
\label{eq6}
\rho(\varphi)=\frac{\dot\varphi^{2}}{2}+V(\varphi)\,,
\end{equation}
and 
\begin{equation}
\label{eq7}
P(\varphi)=\frac{\dot\varphi^{2}}{2}-V(\varphi)\,,
\end{equation}
where $P(\varphi)$ is the inflaton's pressure and $\rho(\varphi)$ is its energy density.
The first term on the right-hand side of Equation (\ref{eq6}) represents the kinetic energy of the field. Inflation occurs when the potential energy density $V(\varphi)$, dominates,  meaning $V(\varphi) \gg {\dot\varphi^{2}}/{2}\,$.
The equation that completes the system is the Klein--Gordon equation for the inflaton 
\cite{InflaQuantumCosmo}:
\begin{equation}
\label{eq8}
\ddot\varphi+3H\dot\varphi=-V^{\prime}(\varphi)\,.
\end{equation}
{The} slow-roll condition is 
$
\label{eq13}
\ddot\varphi \ll 3H\dot\varphi\,.
$
It holds during inflation when the scale factor grows exponentially, and the kinetic energy of the field is negligible~\cite{InflaQuantumCosmo}.

We will apply
these equations and modify the usual quantum field theory estimations for the fluctuations using novel assumptions, as follows.

\section{A First-Order Estimate in \boldmath$V'(\varphi)$}

We consider a mean-field background $\varphi_{c}$,
along with quantum fluctuations, $\delta(\varphi)$,
satisfying $\varphi(t)=\varphi_{c}(t)+\delta(\varphi)$.
In this setup, we assume a decoupled system of equations under the condition that
$
\label{eq89}
\delta\varphi \ll \varphi_{c}(t)
$.
Moreover, the perturbation yields a local change in the expansion rate $H(\varphi)$ proportional to 
\begin{equation}
\label{eq9}
\delta_{H}(k)={\frac{\delta\rho}{\rho}}\propto {-H\delta{t}}\,.
\end{equation}
{Here}, 
we consider that $t \sim {H^{-1}}$, and taking  $\delta{t}=-H^{-2}\,\delta{H}$  {leads us to }$H\delta{t}=-{\delta{H}}/{H}$. Now, considering that $H^{2}\propto{\rho}$, then Equation (\ref{eq4}) leads us to Equation (\ref{eq9}) above. 
Moreover, we can write
{ $\label{eq93}
\delta{t}={{\delta\varphi}/{\dot{\varphi}}}\,.$}
 With the above definitions, Linde (and other authors) obtained~\cite{InflaQuantumCosmo}
\begin{equation}
\label{eq10}
\delta_{H}(k)\sim{\frac{CV^{3/2}(\varphi)}{{M_{pl}}^{3}V^{\prime}(\varphi)}}\,,
\end{equation}
where $C$ is close to ten and $V^{\prime}(\varphi)$ means the first derivative of the potential relative to the scalar field itself.
This quantity is the most essential ingredient of inflation
since it predicts the seeds of posterior structure formation when matter and radiation fall into the 
gravitational wells caused by these perturbations. 
The regions with larger scalar field values inflate rapidly and produce enough density
contrast to trigger structure formation. 

In the case of the massive scalar field with potential $\label{eq20}
V(\phi)={m^{2}{\varphi}^{2}}/{2}\,
$, Equation (\ref{eq10}) yields the following expression for the density contrast produced by  quantum fluctuations: 
\begin{equation}
\label{eq11}
\left(\frac{\delta{\rho}}{\rho}\right)\sim{\frac{Cm{\varphi}^{2}}{M_{pl}^{3}}}\,.
\end{equation}
{This} 
equation and the observational constraint $\delta{\rho}/{\rho}\sim 10^{-5}$ lead to an upper bound for the amplitude of the field: 
%
$\label{eq23}
({\varphi}/{m})^{2} <{10^{-5}}({M_{pl}}/{m})\,.
$
Similar considerations apply to other inflationary models.  

Let us investigate in the next section the expressions that lead us to these formulae, considering the expected expansion of fluctuations in the frequency domain and comparing them with $V^{\prime\prime}$.

\section{Primordial Fluctuations and the Inflationary Potential}
\subsection{Conventional Calculation of Perturbations}
In this section, we examine quantum vacuum fluctuations and, for clarity, repeat the usual derivation of the quantum fluctuations of the scalar field, as we modify this derivation in the next section.
We assume, as usual, that the spacetime is homogeneous and isotropic and described by the FRW equations. 

{Quantum field theory suggests that the quantum vacuum comprises particles and antiparticles of all wavelengths, stretched during inflation to reach astrophysical scales~\cite{Cosmo,Cosmo2}.}

The scalar field, $\varphi$, can be split into 
its unperturbed component 
   (the mean value of the field)
plus a small perturbation, 
$\delta\varphi(\vec{x},t)$, in the form
\begin{equation}
\label{eq12}
\varphi(\vec{x},t)=\varphi_{c}(t)+\delta\varphi(\vec{x},t)\,.
\end{equation}
 {As} the fluctuation amplitude is small ($\delta\varphi<<\varphi_{c}(t)$), the background spacetime is
undisturbed, and the  equation of motion for the field can be decoupled into two equations (inserting  Equation (\ref{eq12}) into Equation (\ref{eq8})):

\begin{equation}
\label{eq99}
\ddot{\varphi_{c}}(t)+3H\dot{\varphi_{c}}(t)=-V^{\prime}(\varphi_{c}(t))\,,
\end{equation}
and
\begin{equation}
\label{eq14}
\delta{\ddot{\varphi}}+3H(\varphi_{c}(t))\delta\dot{\varphi}-{\nabla}^{2}\delta\varphi
=-V^{\prime\prime}(\varphi_{c})\delta{\varphi}\,.
\end{equation}
{The} {first equation describes the evolution of the mean value of the field, while the second equation describes the evolution of the perturbations.
We also assume \mbox{$
\delta\varphi \ll {{H(\varphi_{c})}/{H^{\prime}(\varphi_{c})}}\,.$}} 
 
 In comoving coordinates, the dependence of the wavenumber $k$ on the physical wavenumber is described by 
 \mbox{$k=a(t)\,k_{f}$.} Let us consider the Fourier expansion in normal modes $e^{i\,{k_{f}\, x_{\mu}}}$ for the perturbations.

 If {$k_{f}$} is the physical wavenumber, it is easy to show that the phase of these waves
is invariant under the expansion
\begin{equation}
\label{eq15}
e^{i\,{k\, x_{\mu}/a(t)}}=const\,.
\end{equation}
{Substituting} this into expression Equation (\ref{eq14}), we have a differential
equation for each $k$-mode:
\begin{equation}
\label{eq16}
\delta\ddot{\varphi}_{k}+3H(\varphi)\delta\dot\varphi_{k}+[{(k/a)}^{2}
+V^{\prime\prime}(\varphi)]\delta\varphi_{k}=0\,.
\end{equation}
{First}, we solve  Equation (\ref{eq16}) as usual, with the approximation
${(k/a)}^{2} \gg V^{\prime\prime}(\varphi)$, which is reasonable for short wavelengths. Let us expand the field fluctuation $\delta{\varphi}$
as a sum of creation and destruction operators, 
whose modes are particular solutions of  Equation (\ref{eq16}):
\begin{equation}
\label{eq17}
\delta\varphi=\sum_{k}{\chi_{k}(\eta)A_{k}+\chi_{k}^{*}(\eta)A_{k}^{\dagger}}\,,
\end{equation}
where $A_{k}$ and $A_{k}^{\dagger}$ are the destruction and creation operators, respectively,  $\chi_{k}(\eta)$ are the vibration modes, and $\eta$ is the conformal time. Inserting this solution into Equation (\ref{eq16}), we obtain an equation for the modes,
\begin{equation}
\label{eq18}
\ddot\chi_{k}(t)+3H\chi_{k}(t)+{(k/a)}^{2}\chi_{k}(t)=0\,,
\end{equation}
whose physical solution is given by
\begin{equation}
\label{eq19}
\chi_{k}(\eta)={\frac{a^{-3/2}}{\sqrt{2k/a}}}
\biggl[1+{\frac{iaH}{k}}\biggr]e^{{-ik/aH}}\,.
\end{equation}
{The} expectation value of the $k$-mode's quadratic amplitude is
\begin{equation}
\label{eq92}
<0|{(\delta\varphi)_{k}}^{2}|0>={|\chi_{k}(\eta)|}^{2}\,,
\end{equation}
where
\begin{equation}
\label{eq21}
{|\chi_{k}(\eta)|}^{2}={\frac{H^{2}}{2 \,{k}^{3}}}\biggl[{1+{(k/aH)}^{2}}\biggr]\,.
\end{equation}
{We} obtain the total fluctuation by 
integrating over the total phase space volume:
\begin{equation}
\label{eq22}
<0|{(\delta\varphi)}^{2}|0>={\frac{1}{{(2\pi)}^{3}}}\int{d^{3}k\,
<0|{(\delta\varphi_{k})}^{2}|0>}\,,
\end{equation}
with $d^{3}k = 4\,\pi\,{k}^{2}\,dk$.
Employing Equation (\ref{eq19}), and manipulating some terms, we obtain
\begin{equation}
\label{eq90}
<0|{(\delta{\varphi})}^{2}|0>={\frac{H^{2}}{{4{\pi}^{2}}}}\int_{0}^{Ht} \,(1+{k/H}^{2})\, d\, {\rm ln}(k/H)\,.
\end{equation}
{Evaluating} the integral above and considering the instant when the
perturbation crosses the Hubble horizon as $t={H}^{-1}$, the
integral results in
\begin{equation}
\label{eq24}
<0|{(\delta\varphi)}^{2}|0>={\frac{7\,{H(\varphi)}^{2}}{12\,{\pi}^{2}}}\,.
\end{equation}
{Note} that $<0|(\delta{\varphi})|0>=0$, as can be easily
verified. Taking the square root of this value, we obtain
\begin{equation}
\label{eq25}
\delta\varphi\sim{\frac{H}{2\pi}}\,.
\end{equation}
{Substituting} into Equation (\ref{eq9}), we obtain the
density contrast produced by the quantum fluctuations:
\begin{equation}
\label{eq26}
{\biggl({\frac{\delta\rho}{\rho}}\biggr)}
\sim{\frac{H^{2}(\varphi)}{2\pi\dot{\varphi}_{c}(t)}}\, .
\end{equation}

The results presented in this subsection are the usual textbook treatment of the perturbations.
In the next subsection, we incorporate the term $V^{\prime\prime}(\varphi)$ into the relevant~expressions. 

\subsection{Including Second-Order Variations}
The second-order derivative of the potential $V^{\prime\prime}(\varphi)$ is usually discarded from the calculations in the previous subsection due to the slow-roll conditions~\cite{Cosmo}. 
However, it is straightforward to note that this approximation does not always hold since we cannot ignore the curvature of the potential for any wavenumber $k$. If we ignore this term, then Equation (\ref{eq16}) becomes
$$\delta\ddot{\varphi}_{k}+3H(\varphi)\delta\dot\varphi_{k}+[{(k/a)}^{2}
]\delta\varphi_{k}=0\,.$$

Nevertheless, in Equation (\ref{eq16}), both terms within the brackets must be retained because, as inflation progresses, the first term evolves and is exponentially suppressed with time, making it problematic to neglect the term involving the second derivatives of the potential throughout the entire inflationary period.

Now, we perform the following variable change in Equation (\ref{eq16}):
\begin{equation}
\label{eq27}
k_{eff}(t)=\sqrt{k^{2}+a^{2}(t)V^{\prime\prime}(\varphi_{c})}\,.
\end{equation}
{Using} this equation, the square power of the fluctuations takes the form
\begin{equation}
\label{eq28}
 <0|{(\delta\varphi)}^{2}|0>={\frac{1}{{(2{\pi})}^{3}}}{\int{\frac{{d^{3}k}}{k_{eff}}}\,\biggl[{\frac{1}{2}+\frac{H^{2}}{2\,k_{eff}^{2}}}\biggr]}\,.
\end{equation}
{The} first term in the integral above is the usual fluctuation in the Minkowski 
vacuum (when $H=0$),  while the second term is associated with inflation.
This term arises because the de Sitter space contains particles
$\varphi$ with occupation numbers
\begin{equation}
\label{eq29}
n_{k}={\frac{H^{2}}{2{k}^{2}}}\,.
\end{equation}
{We} see that the occupation numbers will receive contributions from the second-order variations in the
inflationary potentials through the effective wavenumber $k_{eff}(\varphi)$.
The volume element $d^{3}k$ is not modified. Therefore,
if we discard the renormalizable first term, then
Equation (\ref{eq28}) is rewritten as
\begin{equation}
\label{eq30}
<0|{(\delta\varphi)}^{2}|0>={\frac{H^{2}}{4{\pi}^{2}}}
\int{\frac{dk\,{k}^{2}}{{k_{eff}}(\varphi)^{3}}}\,,
\end{equation}
which, in its final form, is
\begin{equation}
\label{eq31}
<0|{(\delta\varphi)}^{2}|0>={\frac{H^{2}}{4{\pi}^{2}}}\int_{0}^{Ht}
{\frac{d\,{\rm ln}(k/H)}{{\left[1+a^{2}(t)V^{\prime\prime}/{k}^{2}\right]}^{3/2}}}\, .
\end{equation}
{In} {this equation, we will consider the following cases: 
$a^{2}(t)V^{\prime\prime}/{k}^{2}\ll {1}$ and  $a^{2}(t)V^{\prime\prime}/{k}^{2}\gg {1}$.}

\subsubsection{First Case: $a^{2}\,V^{\prime\prime}(\varphi_{c})\ll k^{2}$}
Expanding the denominator of Equation (\ref{eq31}) as a power series while taking the first two terms and
considering one period of expansion $t\sim{H^{-1}}$, we obtain
\begin{equation}
\label{eq32}
<0|{(\delta\varphi)}^{2}|0>={\frac{{H}^{2}}{4{\pi}^{2}}}\biggl[{1-{\frac{3(1-{e}^{-2})}{4}}
{\frac{V^{\prime\prime}(\varphi)}{H^{2}(\varphi)}}}\biggr]\,.
\end{equation}
{Taking} the square root power, substituting into Equation (\ref{eq30}), and taking into
account the slow-roll condition,  $\dot{\varphi_{c}}=-{V^{\prime}(\varphi)/3H(\varphi)}$, we have for the density contrast
\begin{equation}
\label{eq33}
\biggl({\delta\rho\over{\rho}}\biggr)\sim{CV^{3/2}(\varphi)
\over{M_{pl}^{3}V^{\prime}(\varphi)}}
{\biggl[1+{3V^{\prime\prime}(\varphi)\over{8H^{2}(\varphi)}}+...\biggr]} \, ,
\end{equation}
{where $C={(3/{2\pi)}}\,{({8\pi/{3}})}^{3/2}\sim{11.6}$ is a normalization constant~\cite{InflaQuantumCosmo}.
The usual \mbox{Equation~(\ref{eq10})}} holds when
$V^{\prime\prime}{(\varphi)}\ll H^{2}(\varphi)$. 

\subsubsection{Second Case: $a^{2}\,V^{\prime\prime}(\varphi_{c})\gg {k}^{2}$}
We write the effective comoving wavenumber as
\begin{equation}
\label{eq34}
k_{eff}=a(t)\sqrt{{(k/a)}^{2}+V^{\prime\prime}(\varphi)}\, .
\end{equation}
{Since} $V^{\prime\prime}(\varphi)$ dominates the first term within
the square root, we obtain
\begin{equation}
\label{eq35}
<0|{(\delta\varphi)}^{2}|0>=
{\frac{H^{2}(\varphi)}{16{\pi}^{3}}}{a(t)}^{-3}
\int{\frac{d^{3}k}{{V^{\prime\prime}(\varphi)}^{3/2}}}\, .
\end{equation}
{The} comoving wavenumber ranges within the interval
\begin{equation}
\label{eq36}
H<k<H{e}^{Ht} 
\end{equation}
because inflation starts at $a=1$ and ends at $a(t)=e^{H t}$.
Therefore, from Equation (\ref{eq35}) we~find
\begin{equation}
\label{eq37}
<0|{(\delta\varphi)}^{2}|0>=
{\biggl({H\over{2\pi}}\biggr)}^{2}{a(t)^{-3}H^{3}\over{3{V^{\prime\prime}(\varphi)}^{3/2}}}
\biggl[e^{3Ht}-1\biggr]\, .
\end{equation}
{Since} $Ht \gg 1$, we obtain the result
\begin{equation}
\label{eq38}
\Delta_{\varphi}=
\sqrt{{<0|{(\delta\varphi)}^{2}|0>}-{<0|(\delta\varphi)|0>}^{2}}
\sim{\frac{H(\varphi)}{2\pi}}\sqrt{\frac{H^{3}(\varphi)}{3{V^{\prime\prime}(\varphi)}^{3/2}}}\,.
\end{equation}
{Therefore}, for very long wavelengths, the fluctuations are corrected according to the density~contrast:
\begin{equation}
\label{eq39}
{\biggl({\frac{\delta{\rho}}{\rho}}\biggr)}_{\lambda>\lambda_{*}}
\sim{\frac{CV^{3/2}(\varphi)}{{M_{pl}}^{3}V^{\prime}(\varphi)}}
\sqrt{H^{3}(\varphi)\over{3{V^{\prime\prime}(\varphi)}^{3/2}}}\,.
\end{equation}
{Next}, we must find the conditions for the wavelength cut-off $\lambda_{c}$.

\subsubsection{Wavelength  Cut-Off}
The comoving $k$ value is constant, while the $k$ cut-off is given by the expression
\begin{equation}
\label{eq40}
a^{2}(t)V^{\prime\prime}(\varphi)={k}^{2}=a^{2}(t){k_{f}}^{2}\,.
\end{equation}
{The} physical wave number is 
$k_{f}={2\pi/{\lambda_{*}}}$. Consequently, the critical wavelength is
\begin{equation}
\label{eq41}
\lambda_{*}(t)={\frac{2\pi}{\sqrt{V^{\prime\prime}(\varphi_{c}(t))}}} \,.
\end{equation}
This equation shows that the $k_{f}$ cut-off depends on  the $V(\varphi)$ profile as on the end time.
\section{Applications}
We will analyze chaotic inflationary models~\cite{linde} under the perspective developed above or~\cite{watson2000,LythLiddle,guth2013}.

\subsection{First Model}
Consider the potential defined by
\begin{equation}
\label{eq42}
V(\varphi)={\frac{m^{2}}{2}}{\varphi}^{2}\,.
\end{equation}
{In} this case, we find the following critical value: 
\begin{equation}
\label{eq43}
\lambda_{*}={\frac{2\pi}{m}}\sim{10^{-32}\,{(m/M_{pl})}^{-1}}\, {\rm cm}\,.
\end{equation}
{Therefore}, unless the mass of the inflaton is negligible in Planck units, almost all the wavelengths enter into the second case (Equation \eqref{eq39}):
\begin{equation}
\label{eq44}
{\biggl({\frac{\delta{\rho}}{\rho}}\biggr)}\sim{CV^{3/2}(\varphi)
\over{{M_{pl}}^{3}V^{\prime}(\varphi)}}
\sqrt{\frac{H^{3}(\varphi)}{3{V^{\prime\prime}(\varphi)}^{3/2}}} \,.
\end{equation}
{Therefore}, the corrected Equation (\ref{eq34}) must be applied. 

If we impose that $${\biggl({\frac{\delta{\rho}}{\rho}}\biggr)}<10^{-5}\,,$$ (the upper bound from observations of the CMB),  it yields a new constraint between the field and mass: 
\begin{equation}
\label{eq45}
{\Gamma\varphi^{7/2}\sim{{m^{2}M_{pl}^{3/2}10^{-5}}}}\,,
\end{equation}
with $\Gamma = \frac{10}{2^{3/2}}\sqrt{\frac{(4\pi/3)^{3/2}}{3}} \sim 5.98$.
Note that this constraint is very different from
\begin{equation}
\label{eq46}
\varphi^{2}\sim{10^{-6}}\frac{M_{pl}^{3}}{m}\,,
\end{equation}
as derived from Equation (\ref{eq10}), from previous analyses. Therefore, our findings constrain the inflaton mass, highlighting its significance. 

\subsection{Second Model}
We now analyze the case with auto-coupling:
\begin{equation}
\label{eq47}
V(\varphi)={\zeta\over{4}}{\varphi}^{4}\,.
\end{equation}
{For} this potential, 
$V^{\prime\prime}(\varphi)=\zeta{\varphi}^{2}$, implying that the field is
time-dependent. 
In this case, Equations (\ref{eq7}) and  (\ref{eq12}) become a coupled system:
\begin{equation}
\label{eq48}
H^{2}=\frac{8\pi\zeta}{12M_{pl}^{2}}\varphi^{4}\,,
\end{equation}
and
\begin{equation}
\label{eq49}
3H\dot{\varphi}=-\zeta{\varphi}^{3}\,.
\end{equation}
{From} these two equations, a first-order differential equation results:
\begin{equation}
\label{eq50}
3D\dot{\varphi}=-\zeta{\varphi}\,,
\end{equation}
with $D=\sqrt{{8 \pi \zeta}/{12 \,M_{pl}^{2}}}$,
whose solution is
\begin{equation}
\label{eq51}
\varphi(t)=\varphi_{i}\,e^{-\frac{\zeta}{3\,D}t}\,.
\end{equation}
{Therefore}, the cut-off evolves in time
as
\begin{equation}
\label{eq52}
\lambda_{*}(t)={\frac{2\pi}{\varphi_{i}\,\sqrt{3\,\zeta}}}\,e^{\frac{\zeta}{3\,D}t}\,.
\end{equation}
{Assuming} a typical value $\zeta\sim{10}^{-12}$~\cite{InflaQuantumCosmo},
  the time scale is $\tau=\frac{1}{M_{pl}}\sqrt{\frac{8\pi}{12\,\zeta}}\sim{10}^{-36}$ {s}
. As time progresses, all wavelengths eventually fall under the first case, making Equation (\ref{eq10}) the accurate expression for the perturbation's amplitude. 

\section{Conclusions}
In this paper, we recalled inflation theory, focusing on its standard formulation and its implications for quantum fluctuations of the scalar field. By incorporating second-order derivatives of the inflationary potential energy into the equation governing these fluctuations, we have extended the theoretical framework to account for more detailed previously overlooked 
corrections.
Equation (\ref{eq39}) describes fluctuations with wavelengths much larger than $\lambda_{*}$, while fluctuations follow Equation (\ref{eq33}) in the near-critical case. Both expressions differ from the usually adopted Equation (\ref{eq10}).

We applied the analysis to chaotic inflation scenarios~\cite{NewInflation,linde} and revealed intriguing
insights.  In the first application, we 
found that the conventional formulae for perturbations
need to be revised 
when considering potentials with the form $m\, \varphi^2 /2 $. In particular, the constraint between the mass and the initial value of the scalar field is very different from other previous analyses~\cite{NewInflation,linde} when we use the amplitude of the fluctuation as an upper~bound.  
    
For potentials like $\zeta\,\varphi^4 /4$, we observed a critical dependence on the duration of the inflationary process. This dependency arises due to the need to compare perturbation modes to a time-dependent critical wavelength, $\lambda_{*}$.
Fluctuations with wavelengths much smaller than the critical value $\lambda_{*}$ exhibit a dependence on the potential and its derivatives, as described by the usual Equation (\ref{eq10}). 
Conversely, fluctuations with wavelengths much larger than $\lambda_{*}$ are characterized by Equation (\ref{eq39}). The predicted fluctuations align with Equation (\ref{eq33}) near criticality.

 {The physical origin of the second-order contributions manifested in \mbox{Equations (\ref{eq33}) and (\ref{eq39})}} can be explained by the presence of the expansion rate, $H$, in those expressions. Indeed, when the subtle variations in the potential ($V''$) are comparable to the Universe expansion rate, then corrections to the conventional expression for the density contrast become relevant. This feature reinforces the importance of the mathematical form chosen for the potential. Physically, it displays a relevant interplay between the rate at which the Universe expands and variations in the changes in the inflationary potential relative to the inflation~field.

Our inclusion of second-order corrections in the quantum fluctuation expression has revealed a significant deviation from the traditional formulations: fluctuations with wavelengths greater than $\lambda_{*}$ will cross the horizon later. Therefore, their implications may be relevant in the distant future of our local Universe. This effect may be responsible for triggering the collapse of our region if its magnitude suffices.  

Equation (\ref{eq10}) is the usual, without corrections from the variations in $V'(\varphi)$. This expression leads us to the known constraints obtained from the comparisons with the CMB, as happens for the upper bound $10^{-5}$ for the amplitude. In the second case investigated in this work, we found deviations in amplitude that implied new constraints. The intermediate case (related to Equation (\ref{eq10})) requires a numerical analysis.

These findings promise to refine our understanding of the inflationary potential and could be fine-tuned using feedback from satellite data. Indeed, this avenue presents an exciting opportunity for future research, which we intend to pursue diligently.



\vspace{6pt}

\section*{Acknowledgments}{{CF
acknowledges FAPESP for support (thematic project \#2013/26258-4). CF thanks CPNq (Brazil) for support (grant \#312454/2021-0).} 
}

 \section*{Conflicts of interest}{{The authors declare no conflicts of interest.} }

\end{document}